\documentclass{article}
\usepackage{spconf,amsmath,graphicx,enumitem, amsfonts}
\usepackage{hyperref}

\title{kNN-CTC: Enhancing ASR via Retrieval of CTC Pseudo Labels}
\name{Jiaming Zhou\textsuperscript{1}, Shiwan Zhao\sthanks{Independent researcher.}, Yaqi Liu\textsuperscript{2},  Wenjia Zeng\textsuperscript{3}, Yong Chen\textsuperscript{3}, Yong Qin\textsuperscript{1}\sthanks{Corresponding author. This work was supported in part by NSF China (Grant No. 62271270).}}
\address
{\textsuperscript{1} Nankai University, Tianjin, China \\
\textsuperscript{2} Beijing University of Technology, Beijing, China\\
\textsuperscript{3} Lingxi (Beijing) Technology Co., Ltd.}
\begin{document}
%
\maketitle
\begin{abstract}
The success of retrieval-augmented language models in various natural language processing (NLP) tasks has been constrained in automatic speech recognition (ASR) applications due to challenges in constructing fine-grained audio-text datastores. This paper presents $k$NN-CTC, a novel approach that overcomes these challenges by leveraging Connectionist Temporal Classification (CTC) pseudo labels to establish frame-level audio-text key-value pairs, circumventing the need for precise ground truth alignments. We further introduce a \emph{``skip-blank''}  strategy, which strategically ignores CTC \emph{blank} frames, to reduce datastore size. 
By incorporating a $k$-nearest neighbors retrieval mechanism into pre-trained CTC ASR systems and leveraging a fine-grained, pruned datastore, $k$NN-CTC consistently achieves substantial improvements in performance under various experimental settings. Our code is available at \href{https://github.com/NKU-HLT/KNN-CTC}{https://github.com/NKU-HLT/KNN-CTC}.

\end{abstract}
\begin{keywords}
speech recognition, CTC, retrieval-augmented method, datastore construction
\end{keywords}
\section{Introduction}
\label{sec:intro}

In recent years, retrieval-augmented language models \cite{knn-lm, RAG, knn-mt,jiang2022towards,alon2022neuro}, which refine a pre-trained language model by linearly interpolating the output word distribution with a $k$-nearest neighbors ($k$NN) model, have achieved remarkable success across a broad spectrum of NLP tasks, encompassing language modeling, question answering, and machine translation. Central to the success of $k$NN language models is the construction of a high-quality key-value datastore.

Despite these advancements in NLP tasks, applications in speech tasks, particularly in automatic speech recognition (ASR), remain constrained due to the challenges associated with constructing a fine-grained datastore for the audio modality.
Early exemplar-based ASR \cite{deselaers2007speech,xu2015exemplar} utilized $k$NN to improve the conventional GMM-HMM or DNN-HMM based approaches.
Recently, Yusuf et al. \cite{text-retrieval-asr} proposed enhancing a transducer-based ASR model by incorporating a retrieval mechanism that searches an external text corpus for potential completions of partial ASR hypotheses. However, this method still falls under the $k$NN language model category, which only enhances the text modality of RNN-T \cite{rnn-t}. Chan et al. \cite{contextual-knn-asr} employed Text To Speech (TTS) to generate audio and used the audio embeddings and semantic text embeddings as key-value pairs to construct a datastore, and then augmented the Conformer \cite{conformer} with $k$NN fusion layers to enhance contextual ASR. However, this approach is restricted to contextual ASR, and the key-value pairs are coarse-grained, with both key and value at the phrase level. Consequently, the challenge of constructing a fine-grained key-value datastore remains a substantial obstacle in the ASR domain.

In addition to ASR, there have been a few works in other speech-related tasks. For instance, RAMP \cite{ramp} incorporated $k$NN into mean opinion score (MOS) prediction by merging the parametric model and the $k$NN-based non-parametric model. Similarly, Wang et al. \cite{xuechen} proposed a speech emotion recognition (SER) framework that utilizes contrastive learning to separate different classes and employs $k$NN during inference to harness improved distances. However, both approaches still rely on the utterance-level audio embeddings as the key, with ground truth labels as the value.

The construction of a fine-grained datastore for the audio modality in ASR tasks is challenged by two significant obstacles: (i) the absence of precise alignment knowledge between audio and transcript characters, which poses a substantial difficulty in acquiring the ground-truth labels (i.e., the values) necessary for the creation of key-value pairs, and (ii) the immense volume of entries generated when processing audio at the frame level. In this study, we present $k$NN-CTC, a novel approach that overcomes these challenges by utilizing CTC (Connectionist Temporal Classification) \cite{ctc} pseudo labels. This innovative method results in significant improvements in ASR task performance. 
By utilizing CTC pseudo labels, we are able to establish frame-level key-value pairs, eliminating the need for precise ground-truth alignments. Additionally, we introduce a \emph{'skip-blank'} strategy that exploits the inherent characteristics of CTC to strategically omit \emph{blank} frames, thereby reducing the size of the datastore. 
$k$NN-CTC attains comparable performance on the pruned datastore, and even surpasses the full-sized datastore in certain instances.
Furthermore, $k$NN-CTC facilitates rapid unsupervised domain adaptation, effectively enhancing the model's performance in target domains.

The main contributions of this work are as follows:

\begin{itemize}[itemsep=2pt,topsep=0pt,parsep=0pt]
\item We introduce $k$NN-CTC, a novel approach that enhances pre-trained CTC-based ASR systems by incorporating a $k$NN model, which retrieves CTC pseudo labels from a fine-grained, pruned datastore.
\item We propose a \emph{``skip-blank''} strategy to reduce the datastore size, thereby optimizing efficiency.
\item We demonstrate the effectiveness of our approach through comprehensive experiments conducted in various settings, with an extension of its application to unsupervised domain adaptation.
\end{itemize}

\section{Our method}
\label{sec:format}

\subsection{$k$NN model}
\begin{figure}[!t]
\centering
\includegraphics[width=0.45\textwidth]{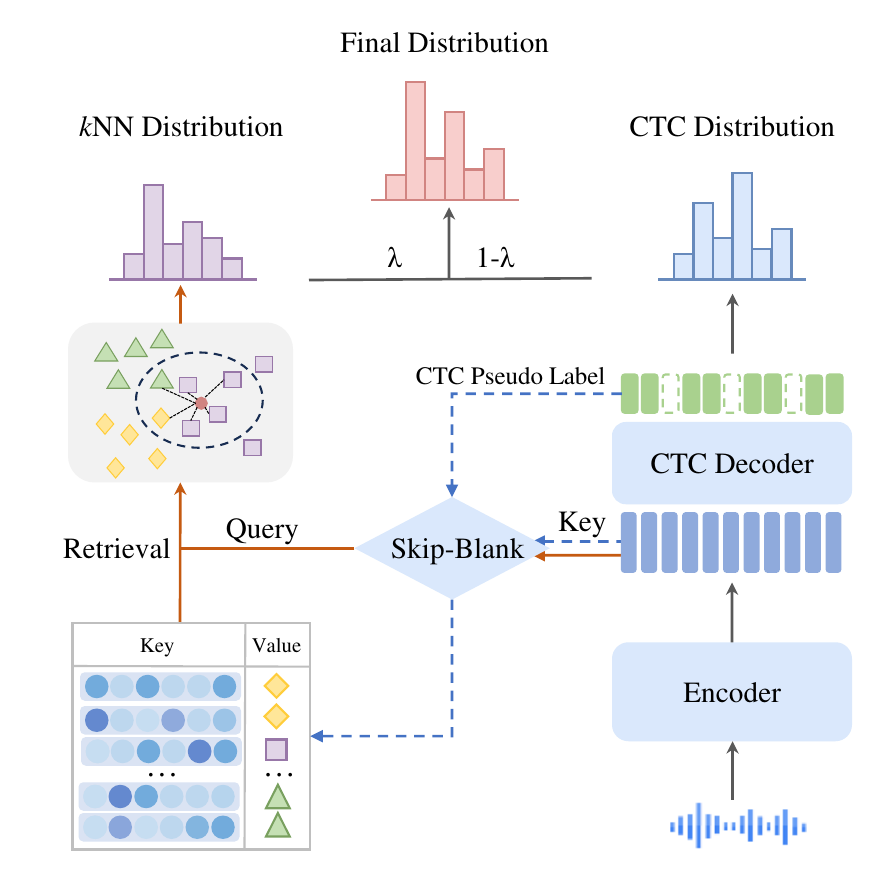}

\caption{Overview of our $k$NN-CTC framework, which combines CTC and $k$NN models. The $k$NN model consists of two stages: datastore construction (blue dashed lines) and candidate retrieval (orange lines).}
\label{overview}
\end{figure}

Figure \ref{overview} presents an overview of our proposed method, which is built upon a CTC-based ASR model and encompasses two stages: datastore construction and candidate retrieval.

\textbf{Datastore construction}: 
Building a fine-grained datastore for the audio modality is challenged by the absence of accurate alignment knowledge between audio frames and transcript characters. This lack of precision complicates the acquisition of ground-truth labels, which are essential for creating audio-text key-value pairs. To tackle this challenge, we adopt the technique from CMatch \cite{cmatch}, utilizing CTC pseudo labels and effectively eliminating the need for precise ground-truth alignments.

With a model trained on labeled data $(X, Y)$, we extract the intermediate representations of $X$, denoted as $f(X)$. 
After evaluating three potential locations, we identify that the input to the final encoder layer's feed-forward network (FFN) yields the most optimal performance, thus selecting it as our keys.
Corresponding values are subsequently obtained.  We derive the CTC pseudo label $\hat{Y}_i$ for the $i$-th frame $X_i$ using the equation:
\begin{equation}
    \hat{Y}_i=arg \mathop{max}\limits_{Y_i}P_{CTC}(Y_i|X_i).
    \label{ctc_loss}
\end{equation}
Thus, we establish frame-level label assignments through the use of CTC pseudo labels. We then designate the intermediate representation $f(X_i)$ as the key $k_i$ and the CTC pseudo label $\hat{Y}_i$ as the value $v_i$, thereby creating an audio-text key-value pair $(k_i, v_i)$ for the $i$-th frame. By extending this process across the entirety of the training set, denoted as $\mathcal{S}$, we construct a datastore $\mathcal{(K,V)}$ composed of frame-level key-value pairs.
\begin{equation}
    (\mathcal{K,V})=\{(f(X_i),\hat{Y}_i)|X_i\in \mathcal{S}\}.
\end{equation}

\textbf{Candidate retrieval}:
During the decoding phase, our process commences by generating the intermediate representation $f(X_i)$ alongside the CTC output $P_{CTC}(Y|X)$. 
Proceeding further, we leverage the intermediate representations $f(X_i)$ as queries, facilitating the retrieval of the $k$-nearest neighbors $\mathcal{N}$. 
We then compute a softmax probability distribution over the neighbors, aggregating the probability mass for each vocabulary item by: 
\begin{equation}
p_{kNN}(y|x) \propto \sum_{(k_i, v_i)\in\mathcal{N},v_i=y} {exp(-d(k_i,f(x)/\tau))},
\end{equation}
where $\tau$ denotes the temperature, $d(\cdot,\cdot)$ signifies the $L^2$ distance. Subsequently, we derive the final distribution $p(y|x)$ by:
\begin{equation}
p(y|x)=\lambda p_{kNN}(y|x)+(1-\lambda)p_{CTC}(y|x),
\end{equation}
where $\lambda$ acts as a hyperparameter, balancing the contributions of $p_{kNN}$ and $p_{CTC}$.
\subsection{Skip-blank strategy}
When processing audio at the frame level, an immense volume of entries is generated, where a considerable portion of the frames are assigned to the \emph{``$<$blank$>$''} symbol due to the characteristic peak behavior of CTC. 
We propose a skip-blank strategy to prune the datastore and accelerate $k$NN retrieval. During datastore construction, this strategy omits frames whose CTC pseudo labels correspond to the \emph{``$<$blank$>$''} symbol, thereby reducing the size of the datastore. This process is indicated by the blue dashed lines in Figure \ref{overview}.
Additionally, during the candidate retrieval stage, $k$NN takes into consideration the CTC decoding output and strategically bypasses frames associated with the \emph{``$<$blank$>$''} symbol, leading to a fast and accurate retrieval. This process is indicated by the orange lines in Figure \ref{overview}.

\label{sssec:subsubhead}
\subsection{Cross-domain adaptation}

Unsupervised domain adaptation (UDA) in ASR \cite{MADI} seeks to enhance the performance within the target domain by leveraging labeled source data $(X_S, Y_S)$ and unlabeled target data $X_T$. 
We develop a rapid unsupervised domain adaptation method for $k$NN-CTC, thereby effectively enhancing its applicability across various domains.
Specifically, given the model trained on labeled source data $(X_S, Y_S)$,  we construct a datastore directly on the unlabeld target data $X_T$, utilizing CTC pseudo labels generated by the model.
During the decoding phase, we employ $k$NN-CTC with the datastore constructed on the unlabeled target data $X_T$ as mentioned above.
Note that our method does not require additional training of the model.

\section{Experimantal setup}
\label{sec:pagestyle}
\subsection{Dataset}
We utilize a diverse set of datasets, including AISHELL-1 \cite{aishell-1} (178 hours, Chinese), Libri-Adapt \cite{libriadapt} (100 hours, English), AISHELL-2 \cite{aishell-2} (1000 hours, Chinese), WenetSpeech \cite{wenetspeech} (10000 hours, Chinese), and Chinese subdialect sets from KeSpeech \cite{kespeech}. These subdialects encompass Mandarin (589 hours), JiangHuai (46 hours), JiLu (59 hours), ZhongYuan (84 hours), and Southwestern (75 hours).
\subsection{Implementation details}
We employ the open-source WeNet \cite{wenet} framework to do all experiments.  All the model's input is log Mel-filter banks with a 25ms window and a 10ms shift. 
Our baseline is the joint CTC-Attention Conformer model.
For AISHELL-1, AISHELL-2, and WenetSpeech we utilize the open-source pre-trained models\footnote{The checkpoints are available at: https://github.com/wenet-e2e/wenet/blob/main/docs/pretrained\_models.en.md}.
For the Libri-Adapt dataset, our ASR model comprises 12 Conformer encoder layers and 6 Transformer decoder layers, each featuring 4 attention heads and 2048 linear units. We set the learning rate to $4\times10^{-4}$, incorporating 25000 warmup steps during training, which spans 150 epochs. The results are derived through CTC greedy decoding. 
During $k$NN-CTC decoding, we utilize FAISS \cite{FAISS} to retrieve the approximate $k$-nearest neighbors where $k$ is set to 1024. $\lambda$ is tuned to get optimal performance on the dev set. $\tau$ is set to 1 during our experiments. For metrics, we employ Word Error Rate (WER) for English datasets and Character Error Rate (CER) for Chinese datasets.

\section{Results}
\label{sec:majhead}
\subsection{In-domain ASR results}
\label{ssec:subhead}

The results of in-domain ASR on three datasets including AISHELL-1 (A-1), AISHELL-2 (A-2), and Libri-Adapt are presented in Table \ref{in-domain-asr}, where the training data is used to construct the datastore. 
Specifically, $k$NN-CTC (pruned) refers to $k$NN-CTC with the skip-blank strategy, whereas $k$NN-CTC (full) denotes standard $k$NN-CTC. 
Clearly, both $k$NN-CTC (full) and $k$NN-CTC (pruned) significantly outperform vanilla CTC on all datasets.
Furthermore, the $k$NN-CTC (pruned), which utilizes the skip-blank strategy, achieves results that are nearly as optimal as the $k$NN-CTC (full), while substantially reducing the size of the datastore.
These patterns suggest that the pervasive occurrence of blank frames compromises the quality of the datastore, negatively impacting retrieval performance.
Leveraging $k$-nearest neighbors retrieval with a fine-grained, pruned datastore, our method improves both performance and efficiency, reducing CER/WER by 5.46\% relatively, and datastore size by 84.56\% impressively.

\begin{table}[t]
\centering
\caption{WER/CER (\%) of in-domain ASR}
\label{in-domain-asr}
\begin{tabular}{ccccc}
\hline
Dataset                & A-1     & A-2     & Libri-Adapt     & Avg.          \\ \hline
CTC             & 5.18          & 6.18          & 12.81          & 8.06          \\ \hline
$k$NN-CTC (full)   & 4.81          & 5.53          & \textbf{12.42} & \textbf{7.59} \\
Datastore size (G)  & 7.12         & 47.12        & 4.82         & 19.69        \\ \hline
$k$NN-CTC (pruned) & \textbf{4.73} & \textbf{5.46} & 12.66          & 7.62          \\
Datastore size (G)  & 0.99         & 6.65         & 1.49          & 3.04         \\ \hline

\end{tabular}
\end{table}

\subsection{Cross-domain ASR results}
\label{ssec:subhead}
To validate the effectiveness of our approach in unsupervised domain adaptation (UDA), we employ the model trained on Wenetspeech (10000 hours) as the source domain. We expand the scope of target domains to include not just the Mandarin Chinese corpus with AISHELL-1 and Mandarin, but also 4 Chinese dialects: JiangHuai, JiLu, ZhongYuan, and Southwestern.
Table \ref{swift-uda} summarizes our experimental results, demonstrating the superior performance of $k$NN-CTC (full) across all tasks compared to CTC, with a relative 4.06\% CER reduction.
Although surpassing CTC, $k$NN-CTC (pruned) falls short of $k$NN-CTC (full). However, the pruned version achieves a substantial 88.57\% datastore size reduction. In the following section, we delve into the ablation study and explore the reasons behind the suboptimal performance of the skip-blank strategy in the UDA scenario.

\subsection{Ablation study}
\label{ssec:subhead}

\begin{figure}[]
\centering
\includegraphics[width=0.45\textwidth]{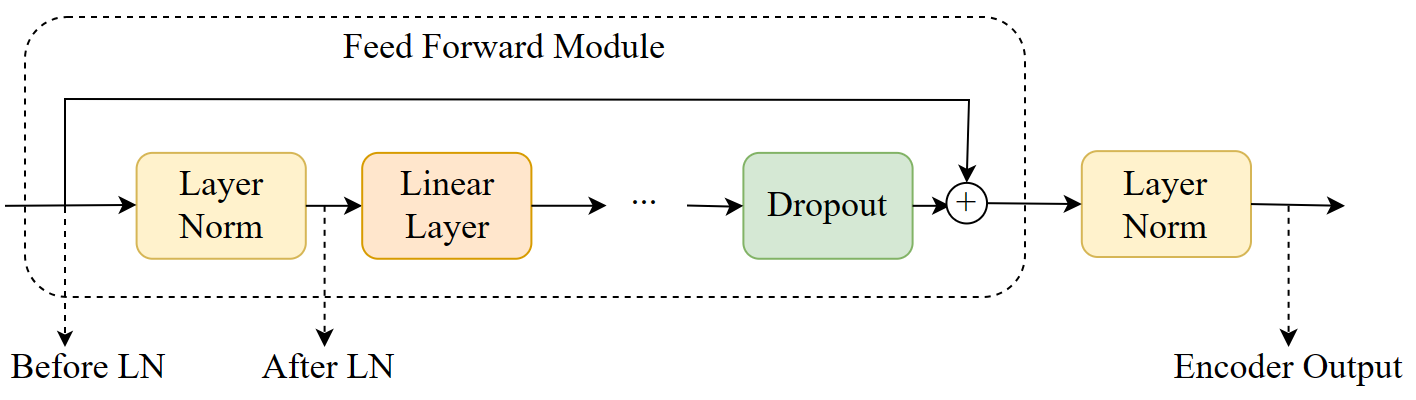}
\caption{Location of keys}
\label{location-of-keys-fig}
\end{figure}

\begin{table*}[]
\centering
\caption{CER (\%) of cross-domain ASR} 
\label{swift-uda}
\begin{tabular}{@{}cccccccc@{}}
\hline
Source Domain  & \multicolumn{6}{c}{Wenetspeech}                                    \\ 
Target Domain  & AISHELL-1 (A-1) & Mandarin & JiangHuai & JiLu  & ZhongYuan & Southwestern &Avg.\\ \hline
CTC                              & 5.02      & 11.13   & 40.87     & 29.76 & 31.94     & 29.2      &24.65   \\ \hline
$k$NN-CTC (full)   & \textbf{4.78}      & \textbf{10.27}   & \textbf{39.95}     & \textbf{28.61} & \textbf{30.04}     & \textbf{28.24}   &\textbf{23.65}     \\
Datastore size (G)                   & 13.33     & 48.38  & 3.88     & 5.05 & 6.99     & 6.34      &14.00  \\ \hline
$k$NN-CTC (pruned)     & 4.83      & 10.95   & 40.34     & 28.86   & 30.53     & 28.41    &23.99    \\
Datastore Size (G)                   & 1.83     & 5.27   & 0.54     & 0.52 & 0.76     & 0.68      &1.60  \\ \hline
\end{tabular}
\end{table*}
\textbf{Location of keys}: To analyze the impact of the locations of keys, we explore three positions in the final Conformer encoder layer shown in Figure \ref{location-of-keys-fig}. As evident in Table \ref{location-of-keys-table}, all key locations improve performance, highlighting the efficacy of our approach. In our experimental assessment, we identify the FFN input after layer normalization as the optimal key location.

\begin{table}[]
\centering
\caption{AISHELL-1 performance with varying key locations }
\label{location-of-keys-table}
\begin{tabular}{cc}
\hline
Key Type                    & CER (\%) \\ \hline
No datastore                & 5.18 \\
Encoder output              & 4.87 \\
FFN input after layer norm  & \textbf{4.73} \\
FFN input before layer norm & 5.17 \\ \hline
\end{tabular}
\end{table}

\begin{figure}[t]
\begin{minipage}[b]{.48\linewidth}
  \centering
  \centerline{\includegraphics[width=4.5cm]{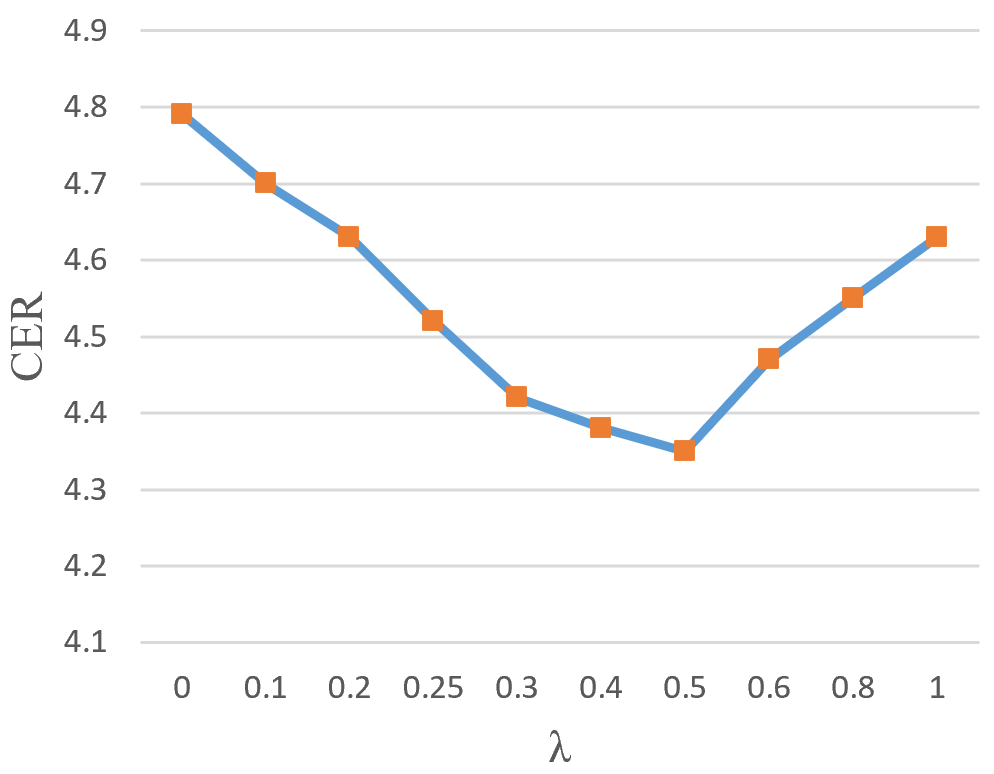}}
  \centerline{(a) A-1 (in-domain)}\medskip
\end{minipage}
\hfill
\begin{minipage}[b]{0.48\linewidth}
  \centering
  \centerline{\includegraphics[width=4.5cm]{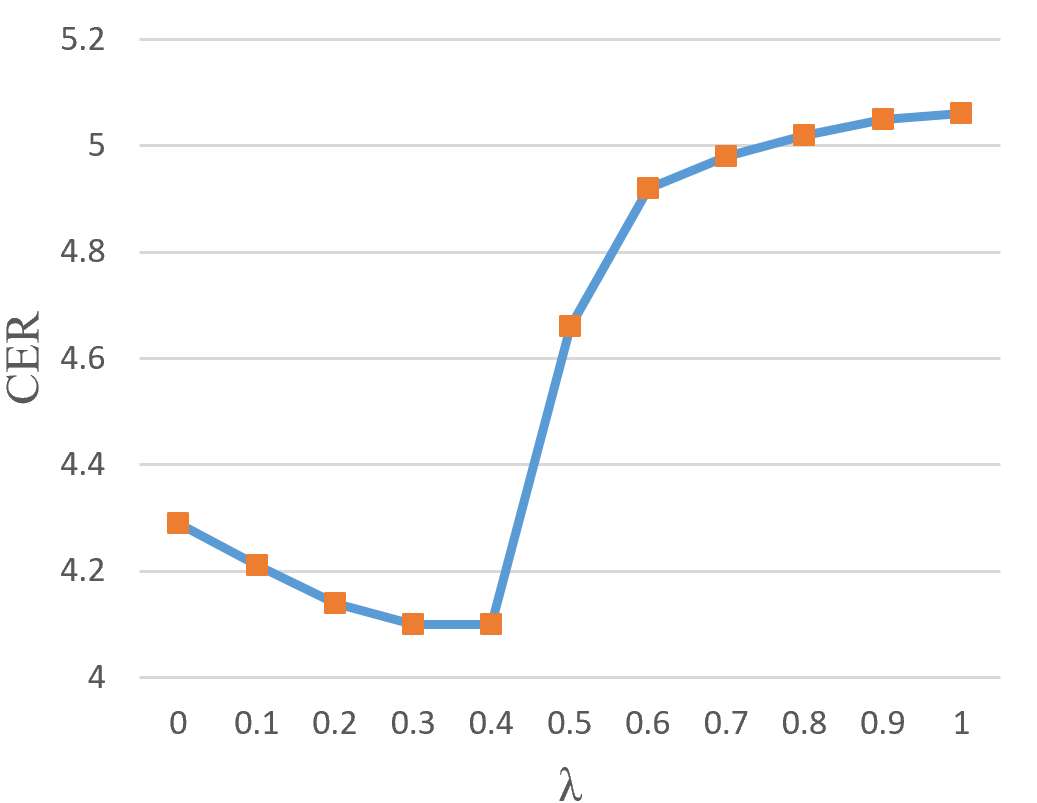}}
  \centerline{(b) W $\to$ A-1 (cross-domain)}\medskip
\end{minipage}
\caption{Effect of hyper-parameter $\lambda$ on DEV set}
\label{tune_lmbda}
\end{figure}

\textbf{Hyper-parameter}: Figure \ref{tune_lmbda} presents $k$NN-CTC (pruned) results of tuning the parameter $\lambda$ on two distinct tasks: in-domain ASR on AISHELL-1 (A-1) and cross-domain ASR from WenetSpeech (W) to AISHELL-1 (A-1).
When $\lambda$ is set to 0, the approach equals to CTC. 
Notably, for in-domain ASR, $k$NN-CTC consistently outperforms CTC when $\lambda>0$.
However, within the cross-domain context, $k$NN-CTC is inferior to CTC for $\lambda>0.4$. 
This discrepancy can be attributed to the inferior quality of pseudo labels of the target data due to the domain shift. This implies that  $k$NN-CTC enhances the ASR systems especially when the pseudo labels utilized for datastore construction are of high quality.

\textbf{The skip-blank strategy}: 
We further investigate the incidence of substitution (S), deletion (D), and insertion (I) errors within the CTC model. Analysis reveals substitution errors (S) are most prevalent. The $k$NN approach corrects S errors by substituting incorrect non-blank symbols with correct ones. Similarly, it corrects D errors by replacing blank symbols with non-blanks, and I errors by substituting non-blanks with blanks.

$k$NN-CTC (pruned) is limited to addressing only S errors.
As shown in Table \ref{comparison}, the percentage of S errors decreases in the cross-domain scenario. While $k$NN-CTC (full) reduces I errors, leading to a better performance than $k$NN-CTC (pruned), it can potentially increase D errors due to the pervasive occurrence of blank frames in the datastore.
Nevertheless, $k$NN-CTC (pruned) achieves similar performance to $k$NN-CTC (full) but with substantially reduced datastore size. This highlights the pruned version's efficiency and efficacy in managing CTC errors despite tradeoffs. This study thus lays a promising groundwork for further optimizing the $k$NN-CTC framework, potentially paving the way for even more streamlined and efficient models in the future.

\begin{table}[]
\centering
\caption{Quantities of Substitution (S), Deletion (D), and Insertion (I) errors}
\label{comparison}
\begin{tabular}{@{}cccccc@{}}
\hline
Task      & Method       & S    & D   & I    & CER (\%)  \\ \hline
A-1 & CTC         & 5185 & 140 & 97   & 5.18  \\
A-1 & +$k$NN (full)  & 4715 & 252 & 72   & 4.81  \\ 
A-1 & +$k$NN (pruned) & 4717 & 139 & 96   &\textbf{4.73}  \\ \hline
W$\to$A-1       & CTC         & 4827 & 148 & 278 & 5.02 \\
W$\to$A-1       & +$k$NN (full)  & 4625 & 154 & 225 & \textbf{4.78} \\ 
W$\to$A-1       & +$k$NN (pruned) & 4632 & 148 & 278 & 4.83 \\
\hline
\end{tabular}
\end{table}

\section{Conclusion}
\label{sec:print}

In this paper, we propose $k$NN-CTC to enhance 
CTC-based ASR systems with the retrieval mechanism. Leveraging CTC pseudo labels, we construct a fine-grained audio-text datastore. We further introduce the skip-blank strategy to reduce the datastore size. Comprehensive experiments demonstrate the effectiveness of $k$NN-CTC in enhancing ASR systems in both in-domain and cross-domain scenarios.

\newpage
\bibliographystyle{IEEEbib}
\bibliography{strings,refs}

\end{document}